\documentclass[a4paper,12pt]{article}
\title{The Kerr medium as an {\sf SU(2)} system}
\author{K V S Shiv Chaitanya$^1$\footnote{chaitanya@bits-hyderabad.ac.in}, 
~~Prasanta K Panigrahi$^2$\footnote{pprasanta@iiserkol.ac.in}, \\
V Srinivasan$^3$\footnote{vsspster@gmail.com} ~~and ~~A S Vytheeswaran$^3$\footnote{vythee@imsc.res.in; ~asvythee@gmail.com}\\
{}\\
$^1$\textit{BITS Pilani, Hyderabad Campus, Jawahar Nagar}, \\\textit{Shameerpet Mandal, Hyderabad, India 500 078.}\\
$^2$\textit{Indian Institute of Science Education and Research (IISER)},\\ \textit{Kolkata, Mohanpur Campus, Nadia, India, 714 252.}\\$^3$\textit{Department of Theoretical Physics, University of Madras,}\\  \textit{Guindy Campus, Chennai, India, 600 025.}}

\date{}
\begin{document}

\maketitle

\begin{abstract}
The Kerr medium in the presence of damping and associated with SU(1,1)
symmetry, is solved using the techniques of Thermo field Dynamics (TFD).
 These TFD techniques, well studied earlier \cite{vs3},
help us to exactly solve the Kerr medium as a spin damped system
associated with SU(2) symmetry. Using TFD, the association with SU(2) 
is exploited to express the dynamics of the system as a Schrodinger-like 
equation, whose solution is obtained using the appropriate disentanglement
theorem. These considerations are extended to a system with multi-mode coupled
nonlinear oscillators."

\end{abstract}

\section{Introduction}

	 In nonlinear optics, one of the most important processes is propagation in the Kerr medium. This medium is used extensively for the generation of nonclassical light, for example, squeezed light \cite{sqe} which is very commonly used in quantum optics. Another type of nonclassical light generated by this 
medium  are the cat states known as the Yurke-Stoler states \cite{cat}. 

While considering various aspects of the Kerr medium, a natural thing that arises is  the presence and effect of damping. The dynamics for this system has been studied and solved in \cite{vs3}. In this paper, the authors have also 
solved the problem of the Kerr medium with {\sf N} coupled oscillators in the presence of damping \cite{vs4}.  In both these cases the method of thermo-field dynamics is used for solving the master equation. 
 
The method of thermo field dynamics \cite{U1,U2,oj,U3,U4} is developed in
finite temperature field theory. 
This formalism has two salient features. First, the solving of the master equation is reduced to solving a Schr{\" o}edinger-type equation. Thus all the techniques available for solving the
Schroedinger equation are applicable here. Second, the thermal coherent state under the master equation evolution goes over to a thermal coherent state. Thus, the application of this method
in solving the master equation  gives a very simple and elegant approach. 

\vspace{2mm}
A brief description of thermofield dynamics (TFD) is given below. In TFD the state $\vert 
\rho^\alpha\rangle,  1/2  \le \alpha \le 1$, is considered a state vector in the extended 
Hilbert space ${\cal H}\otimes {\cal H}^*$. Then the average of any operator ${\hat{A}}$ 
with respect to $\rho$ reduces to a scalar product,
\begin{equation}
\langle A\rangle = {\sf Tr}(A\rho) = \langle\rho^{1-\alpha}\vert A\vert \rho^\alpha\rangle,
\end{equation}
where the state $\vert \rho^\alpha\rangle$ is given by
\begin{equation}
\vert \rho^\alpha\rangle = \hat{\rho}^\alpha\vert I\rangle,
\end{equation}
where
\begin{equation}
\hat{\rho}^\alpha = \rho^\alpha \otimes I,
\end{equation}
and $\vert I\rangle$ is the resolution of the
identity
\begin{equation}
\vert I\rangle=\sum \vert N\rangle\langle N\vert = \sum \vert N\rangle\otimes\vert \tilde{N}\rangle \equiv \sum \vert N,\tilde{N}\rangle,
\end{equation}
in terms of a complete orthonormal set  $\{\vert N\rangle\}$  in  ${\cal  H}$.
The state vector $\vert I\rangle$ takes a normalized vector to another
normalized vector in the extended Hilbert space ${\cal H}\otimes {\cal 
H}^*$.
 From now on, we work in,
$\alpha=1$ representation\cite{vs3,vs4,kap}.
In this representation, for any hermitian operator $A$, one has 
\begin{equation}
\langle A\rangle = Tr(A\rho) = \langle A\vert\rho\rangle.
\end{equation} 
By choosing the number state $\vert n\rangle$ for  $\vert N\rangle$
we introduce the creation and  annihilation 
operators $a^\dagger, \tilde{a}^\dagger,  a$,  and  $\tilde{a}$ and their actions 
as follows
\begin{eqnarray}
a\vert n,m\rangle &=& \sqrt{n} \vert n-1,m\rangle, 
~~~~~~a^\dagger\vert n,m\rangle = \sqrt{n+1} \vert n+1,m\rangle,\\
\tilde{a}\vert n,m\rangle &=&  
\sqrt{m} \vert n,m-1\rangle, ~~~~~\tilde{a}^\dagger\vert n,m\rangle 
 = \sqrt{m+1} \vert n,m+1\rangle.
\end{eqnarray}
Further, the operators $a$ and $a^\dagger$  commute  with  $\tilde{a}$  and 
$\tilde{a}^\dagger$.  It is clear from the above that
$a$ acts on the vector space $\cal{H}$ and $\tilde{a}$ acts on vector space $\cal{H^*}$. From the 
expression for $\vert I\rangle$ in terms of the number states
\begin{equation}
\vert I\rangle = \sum_{n,\tilde{n}} \vert n,\tilde{n}\rangle,
\end{equation}
it follows that
\begin{equation}
a\vert I\rangle=\tilde{a}^\dagger \vert I\rangle,\; a^\dagger\vert I\rangle = 
\tilde{a}\vert I\rangle,\label{10}
\end{equation}
and hence for any operator one has
\begin{equation}
A\vert I\rangle = \tilde{A}^\dagger \vert I\rangle,
\end{equation}
where $\tilde{A}$ is obtained from $A$ by making the  replacements 
(called tilde  conjugation  rules)  $a\to   \tilde{a},   a^\dagger   \to 
\tilde{a}^\dagger,  $ along with complex numbers going to their conjugates. For a more detailed discussion on TFD
see ref\cite{U1,U2,U3,U4}.

\subsection{The Master Equation}
 
Given any master equation of the form
\begin{equation}
 \frac{\partial}{\partial t}\rho(t) ~= ~-\frac{i}{\hbar}(H\rho-\rho H)+L\rho,
\end{equation}
one converts this into a problem in TFD by applying $\vert I\rangle$ from the right to get
\begin{equation}
\frac{\partial}{\partial t}\vert \rho(t)\rangle = -i\hat{H}\vert \rho\rangle\label{sc}
\end{equation}
where
\begin{equation}
 -i\hat{H}=i(H-\tilde{H})+L.
\end{equation}
In TFD $-i\hat{H}$ is {\it tildian} i.e. it reflects the hermiticity property of the density 
operator. Thus the problem of solving the master equation is reduced to solving a Schr{\" 
o}edinger equation. 

As illustration, consider the master equation for a damped linear harmonic oscillator. 
In TFD \cite{vs3,vs4,kap,vs1} this is given by

\begin{eqnarray}
 \frac{\partial}{\partial t}\vert \rho(t)\rangle &=&\Big[\left(\frac{\kappa(\bar{n}+1)}{2}
\left(2a\tilde{a}-a^{\dagger}a-\tilde{a} \tilde{a}^{\dagger}\right)
+\frac{\kappa\bar{n}}{2}
\left(2a^{\dagger}\tilde{a}^\dagger-aa^{\dagger}- \tilde{a}^{\dagger}\tilde{a}\right)\right)
\nonumber\\&& ~~~~- i\omega(a^{\dagger}a-\tilde{a} \tilde{a}^{\dagger})\Big]\vert \rho(t)\rangle.\label{mas},
\end{eqnarray}
where $ \kappa, ~{\bar{n}} $ ~are as defined in \cite{vs3,vs4,kap,vs1}.
This model of the master equation is considered for many dissipative systems.  

We define the following operators
\begin{eqnarray}
\mathcal{K}_3 & = & \frac{1}{2}\left(a^\dagger a +\tilde{a}^{\dagger}\tilde{a} + 1\right), ~~~~\mathcal{K}_+=a^\dagger \tilde{a}^\dagger, 
~~~~\mathcal{K}_-=a \tilde{a}
\end{eqnarray}
and note that these  satisfy the ${\sf SU}(1,1)$ algebra
\begin{eqnarray}
[\mathcal{K}_3,\mathcal{K}_+ ]&=&\mathcal{K}_+, ~~~~~~[\mathcal{K}_3,\mathcal{K}_- ]=\mathcal{K}_-, 
~~~~~~[\mathcal{K}_+,\mathcal{K}_- ]=2\mathcal{K}_3,
\end{eqnarray}
The operator ~ $\mathcal{K}_- = \left(a^{\dagger}a - \tilde{a}^{\dagger}\tilde{a}\right)$ ~is here the Casimir operator. 

In terms of these ${\sf SU}(1,1)$  generators the master equation (\ref{mas}) becomes
\begin{eqnarray}
 \frac{\partial}{\partial t}\vert \rho(t)\rangle &=& -i\hat{H}\vert \rho\rangle\nonumber\\
 & = & \left(-i\omega\mathcal{K}_0+\kappa(\bar{n}+1)\mathcal{K}_-
 +\kappa\bar{n}\mathcal{K}_+ -\kappa(2\bar{n}+1)\mathcal{K}_3 +\frac{\kappa\bar{n}}{2}
\right)\vert \rho(t)\rangle.\label{masy}
\end{eqnarray}
It is clear that the master equation above is like a Schr{\" o}edinger equation  
associated with ${\sf SU}(1,1)$ symmetry. Its solution is given by the evolution operator
~$ e^{-i\hat{H}t} $, with the $ \hat{H} $ as given above. 
One can then apply the disentanglement theorem (for {\sf SU(1,1)}) and solve for arbitrary initial conditions.  

\vspace{4mm}
In this paper, we look at systems for which {\sf SU(2)} operators (generators) can be introduced;
thus these systems will be associated with {\sf SU(2)}  symmetry. The TFD method is used, and
the Shr{\" o}edinger type of equation thus obtained is considered using the disentanglement theorem for 
{\sf SU(2)}.

\section{The Kerr Medium}
In this section, we consider the  Kerr medium in presence of 
damping as a spin system. The Hamiltonian for the Kerr medium is 
\begin{equation}
H=\omega  a^\dagger  a+\chi(a^\dagger a)^2
\end{equation}
where $\omega$ is the frequency of the oscillator, and $\chi$ is the coupling factor which depends on the Kerr
medium.

The evolution of this Hamiltonian is given by the master equation
\begin{equation}
{\partial\over\partial t}     \rho     =     -i[H, \rho]     \label{mdr}
\end{equation}
Applying $\vert I\rangle$ on  (\ref{mdr})  from  the 
right and using (\ref{10}) this master equation for $\rho$ goes over to a 
Schr\"odinger-like equation for the state $\vert \rho\rangle$ 
\begin{equation}
{\partial\over\partial t} \vert\rho\rangle  = -i\hat{H}\vert\rho\rangle,
\end{equation}
where
\begin{eqnarray}
-i\hat{H} & = & - i\omega(a^\dagger a-\tilde{a}^\dagger\tilde{a}) 
-i\chi [(a^\dagger a)^2-(\tilde{a}^\dagger\tilde{a})^2]  
\end{eqnarray}
This Hamiltonian can be rewritten as
\begin{eqnarray}
-i\hat{H} & = & -i\omega(a^\dagger a-\tilde{a}^\dagger\tilde{a}) 
-i\chi [(a^\dagger a+\tilde{a}^\dagger\tilde{a})(a^\dagger a-\tilde{a}^\dagger\tilde{a})] \label{hj} 
\end{eqnarray}
Then one can identify this Hamiltonian (\ref{hj}) with an ${\sf SU}(1,1)$ system
\begin{eqnarray}
-i\hat{H}  &=&-i\omega\mathcal{K}_0
- 2i\chi \mathcal{K}_0\left(\mathcal{K}_3 - \frac{1}{2}\right). \label{hj1} \\
\mathcal{K}_0 & = & \rule{0mm}{10mm}(a^\dagger a-\tilde{a}^\dagger\tilde{a}), ~~~~~\mathcal{K}_3 = 
\frac{1}{2}\left(a^\dagger a  + \tilde{a}^\dagger\tilde{a} + 1\right) \nonumber
\end{eqnarray}
with $ \mathcal{K}_0 $ the Casimir operator.
The formalism of thermofield dynamics thus confirms the ${\sf SU}(1,1)$ nature of the
non-linear oscillator, with the above $ \mathcal{K}_0 $ as the Casimir operator, and the $ \mathcal{K}_3 $
generator as given above. 
This ${\sf SU}(1,1)$ system has been studied in detailed in \cite{vs3,vs4}. 

\vspace{1mm}
It is interesting that when the roles of $ \mathcal{K}_0 $ and $ \mathcal{K}_3 $
are (somewhat) reversed, we have an identification with ${\sf SU}(2)$, implying 
that the Kerr medium can be viewed as a {\sf SU(2)} system.
To see this we define the following operators
\begin{eqnarray}
\mathcal{S}_0 = a^\dagger a + \tilde{a}^{\dagger}\tilde{a}, ~~~~~\mathcal{S}_3 & = & \frac{\left(a^\dagger a - \tilde{a}^{\dagger}\tilde{a}\right)}{2}, ~~~~~\mathcal{S}_+=a^\dagger \tilde{a}, ~~~~~\mathcal{S}_-=a \tilde{a}^\dagger.
\end{eqnarray}
These satisfy the {\sf SU(2)} algebra
\begin{eqnarray}
[\mathcal{S}_3,\mathcal{S}_+]& = & \mathcal{S}_+, ~~~~~[\mathcal{S}_3,\mathcal{S}_- ]= -\mathcal{S}_-, 
~~~~~[\mathcal{S}_+,\mathcal{S}_- ] = 2\mathcal{S}_3,
\end{eqnarray}
with ~$ \mathcal{S}_0 = \left(a^\dagger a +\tilde{a}^{\dagger}\tilde{a}\right) $ ~as the Casimir operator.
The Hamiltonian (\ref{hj}) for the Kerr medium is thus identified with having {\sf SU(2)} symmetry. In 
terms of these generators, we have 
\begin{eqnarray}
\hat{H} & = & \omega\mathcal{S}_3
+ \chi \mathcal{S}_0\mathcal{S}_3.  \label{hj2} 
\end{eqnarray}
 Hence, by adding a spin damping to this Hamiltonian (\ref{hj2}) one has 
\begin{eqnarray}
\hat{H}_{\sf D} &=& \omega\mathcal{S}_3 + \chi\mathcal{S}_0\mathcal{S}_3 + i\gamma\mathcal{S}_+ + i\gamma \mathcal{S}_- -   i\gamma\mathcal{S}_3 \label{spo}
\end{eqnarray}
where $\gamma$ is the decay parameter of the dissipative cavity. This, when substituted in the Schr{\" o}dinger
equation in (20) results in its solution, which is written as
\begin{equation}
\vert \rho(t)\rangle =\exp(\gamma_+\mathcal{S}_+ 
 +\gamma_3\mathcal{S}_3 + \gamma_-\mathcal{S}_-)\vert \rho(0)\rangle,
\end{equation}
where 
\begin{equation}
\gamma_+ = \gamma t; ~~~~\gamma_-=\gamma t; ~~~~\gamma_3 
= -(i\omega +\gamma + i\chi\mathcal{S}_0)t. 
\end{equation}
Using the disentangling theorem \cite{bch, bch1}, one has
\begin{equation}
\vert \rho(t)\rangle ~= ~\exp\left(\Gamma_+\mathcal{S}_+\right) 
\exp\left(\Gamma_3\mathcal{S}_3\right) \exp\left(\Gamma_-\mathcal{S}_-\right)\mid\rho(0)\rangle , \label{hui}
\end{equation}
where
\begin{eqnarray}
\Gamma_{\pm} & = & \left(\frac{\sinh\lambda}{\lambda\cosh\lambda - \gamma_3\sinh\lambda}\right)
\gamma_{\pm}; ~~~~~~~~\Gamma_3 = \ln\left(\frac{\lambda}{\lambda\cosh\lambda - \gamma_3\sinh\lambda}\right)
\end{eqnarray}
with
\begin{eqnarray}
\lambda^2 = \gamma^2_3 + \gamma_+\gamma_-
\end{eqnarray}

\vspace{1mm}
To go further, we expand the state  $\vert \rho(0)\rangle$  in the basis of number 
states,
\begin{eqnarray}
\vert \rho(0)\rangle=\sum_{m,n = 0}^{\infty}\;\rho_{m,n}(0)\vert m,n\rangle\label{rio1}
\end{eqnarray} 
We substitute this expansion in (30) and use the explicit expressions for 
$ \mathcal{S}_{\pm}, \mathcal{S}_3 $. Taking care that the repeated 
actions of  ~$ a, \tilde{a} $ ~on the number states $ \vert m,n\rangle $ 
is limited by the values of $ m, n $, we get the result

\begin{eqnarray}
\vert \rho(t)\rangle ~& = & ~\sum_{m,n = 0}^{\infty}\;\rho_{m,n}(0)\;\sum_{r=0}^m\frac{(\Gamma_{-})^{r}}{r!}\sum_{s=0}^{(n+r)}\frac{(\Gamma_{+})^{s}}{s!}\;\exp\left[{\displaystyle{\Gamma_3\frac{(m-n-2r)}{2}}}\right]\\
& & \rule{0mm}{8mm} ~~~~~~\times \sqrt{(m)_r\;(n+1)^{(r)}(n+r)_s(m-r+1)^{(s)}} \;\;\;\vert(m-r+s), (n+r-s)\rangle\nonumber
\end{eqnarray}

\vspace{1mm}
where we have used the Pochhammer notation \cite{poch}, defined by $(J)_n=J(J-1)(J-2)\cdots (J-n+1)$ and $(J)^{(n)} = J(J+1)(J+2)\cdots (J+n-1)$.
~This completes the solution.

\section{The Master Equation for Coupled Non-Linear Oscillators}

We generalize the above work for $N$ coupled oscillators in the presence 
of a Kerr medium. The dynamics is given by  
\begin{eqnarray}
{\partial\over\partial t}     \rho     &=&     -i\sum_{i=1}^N\omega_i [ a_i^\dagger  a_i,\rho]-i\sum_{i,j=1}^N\chi_{ij}\;\left[(a_i^\dagger a_i) (a_j^\dagger a_j), ~\rho\right], \label{mdrm}
\end{eqnarray}
where the ~$ \chi_{ij} $ ~indicates the coupling among the oscillators and 
may depend on the Kerr medium.
We take ~$ \chi_{ij} = \chi_{ji} $. 
~Applying $\vert I\rangle$ on  (\ref{mdrm})  from  the 
right and using (\ref{10}) this master equation for $\rho$ goes over to a 
Schr\"odinger-like equation for the state $\vert\rho\rangle$ 
\begin{equation}
{\partial\over\partial t} \mid \rho\rangle  = -i\hat{H}\mid \rho\rangle\,\,\,,
\end{equation}
where
\begin{eqnarray}
-i\hat{H} &=& -i\sum_{i=1}^N\omega_i \left(a_i^\dagger a_i-\tilde{a}_i^\dagger\tilde{a}_i\right) 
-i\sum_{i,j=1}^N\chi_{ij}\;\left[a_i^\dagger a_ia_j^\dagger a_j-\tilde{a}_i^\dagger\tilde{a}_i\tilde{a}_j^\dagger\tilde{a}_j\right].  
\end{eqnarray}
This Hamiltonian can be rewritten as
\begin{eqnarray}
-i\hat{H} &=&-i\sum_{i=1}^N\omega_i (a_i^\dagger a_i-\tilde{a}_i^\dagger\tilde{a}_i) 
-i\sum_{i,j=1}^N\chi_{ij}\;(a_i^\dagger a_i+\tilde{a}_i^\dagger\tilde{a}_i)(a_j^\dagger a_j-\tilde{a}_j^\dagger\tilde{a}_j)
\end{eqnarray}
It has been shown in \cite{vs4} that this Hamiltonian is related to {\sf SU(1,1)}, with appropriate definitions of
the group generators. The case of damped coupled oscillators has also been considered, and the solution to the
evolution equation obtained using the disentanglement theorem for {\sf SU(1,1)}.

Here we study the relation between this system and {\sf SU(2)}. By including damping to this Hamiltonian one has
\begin{equation}
-i\hat{H}_D     =  \sum_{i=1}^N\left[-i\omega_i\mathcal{S}^i_3 - i\sum_{j=1}^N\chi_{ij}\mathcal{S}^i_0\mathcal{S}^i_3 + \gamma_i\mathcal{S}^i_-
+\gamma_i 
\mathcal{S}^i_+   -   \gamma_i \mathcal{S}^i_3\right],\label{jio}
\end{equation}
where $\gamma$ is the decay parameter for the dissipative cavity. The operators
\begin{eqnarray}
\mathcal{S}_3^i & = &\frac{(a_i^\dagger a_i-\tilde{a}_i^\dagger\tilde{a}_i)}{2}, ~~~~~\mathcal{S}_+^i=a_i^\dagger \tilde{a}_i, ~~~~~\mathcal{S}^i_-=a_i \tilde{a}_i^\dagger
\end{eqnarray}
satisfy the {\sf SU(2)} algebra
\begin{eqnarray}
\left[\mathcal{S}^i_3,\mathcal{S}^i_+\right] & = & \mathcal{S}^i_+, ~~~~~\left[\mathcal{S}^i_3,\mathcal{S}^i_-\right]  = -\mathcal{S}^i_-, ~~~~~\left[\mathcal{S}^i_+,\mathcal{S}^i_-\right] = 2\mathcal{S}^i_3,
\end{eqnarray}
with ~$\mathcal{S}_0^i = \left(a_i^\dagger a_i+\tilde{a}_i^\dagger\tilde{a}_i\right)$ ~the Casimir operator for
{\sf SU(2)}.

\vspace{1mm}
The substitution of the Hamiltonian  (\ref{jio})  in (36) leads to the solution
 of (36),
\begin{equation}
\vert \rho(t)\rangle =\prod_{i=1}^N  \exp\left(\gamma^i_+\mathcal{S}^i_+ 
 +\gamma_3^i\mathcal{S}^i_3 + \gamma^i_-\mathcal{S}^i_-\right)\;\vert \rho(0)\rangle,
\end{equation}
where 
\begin{equation}
\gamma_+^i = \gamma_i t; ~~~~~\gamma^i_-=\gamma_i t;  ~~~~~\gamma^i_3 
= -\left(i\omega_i + \gamma_i + i\sum_{j=1}^N\chi_{ij}\mathcal{S}^j_0\right)t.
\end{equation}
Using the disentangling theorem, one has
\begin{equation}
\vert \rho(t)\rangle = \prod_{i=1}^N \exp\left(\Gamma^i_+\mathcal{S}^i_+\right) 
\exp\left(\Gamma^i_3\mathcal{S}^i_3\right) \exp\left(\Gamma^i_-\mathcal{S}^i_-\right)\;\vert\rho(0)\rangle,
\label{huio}
\end{equation}
where
\begin{eqnarray}
\Gamma_{i\pm}&=&\left(\frac{\sinh\phi_i}{\phi_i \cosh\phi_i - \gamma_{i3}\sinh\phi_i}\right)\gamma_{i\pm}, 
~~~~~\Gamma_{i3}=\ln\left(\frac{\phi_i}{\phi_i\cosh\phi_i - \gamma_{i3}\sinh\phi_i}\right)
\end{eqnarray}
with
\begin{eqnarray}
\phi_i^2 & = & \gamma_{i3}^2
 + \gamma_{i+}\gamma_{i-}.
\end{eqnarray}
Thus the evolution of the state for the coupled nonlinear oscillator system 
is given in terms of the generators of {\sf SU(2)}.

As earlier, we make an expansion in the basis of number states. We first introduce 
$$
{\bf m} = (m_1, m_2, m_3, \ldots, m_{\sf N})\hspace{1.5cm}{\bf n} = (n_1, n_2, n_3, \ldots, n_{\sf N}).
$$
Then we make the expansion for $\vert \rho(0)\rangle$ as
\begin{eqnarray}
\vert \rho(0)\rangle=\sum_{{\bf m}, {\bf n}=0}^{\infty}\rho_{{\bf m}, {\bf n}}(0)\vert {\bf m},{\bf n}\rangle.\label{rio2}
\end{eqnarray}
It is to be understood that the sum above is over all the $m_i$ and the $n_i, \;\;(i = 
1,2,3\ldots, {\sf N}) $. 

Using this in  (\ref{huio}) one gets
\begin{eqnarray}
\vert \rho(t)\rangle ~& = & ~\prod_{i=0}^N\;\sum_{m_i, n_i=0}^{\infty}\;\rho_{m_i,n_i}(0)\;\sum_{r_i=0}^{m_i}\frac{(\Gamma_{i-})^{r_i}}{r_i!}\sum_{s=0}^{(n_i+r_i)}\frac{(\Gamma_{i+})^{s_i}}{s_i!}\;\exp\left[{\displaystyle{\Gamma_{i3}\frac{(m_i-n_i-2r_i)}{2}}}\right]\\
& & \rule{0mm}{8mm}\times \sqrt{(m_i)_{r_i}\;(n_i+1)^{(r_i)}(n_i+r_i)_{s_i}(m_i-r_i+1)^{(s_i)}} \;\;\vert(m_i-r_i+s_i), n_i+r_i-s_i)\rangle\nonumber
\end{eqnarray}
here $(J)_n=J(J-1)(J-2)\cdots (J-n+1)$ and $(J)^n=J(J+1)(J+2)\cdots (J+n-1)$ is the Pochhammer's notation.
This completes the solution.

\section{Conclusion}
In this paper, using techniques from thermofield dynamics we have considered the Kerr medium as an {\sf SU(2)} system with damping. This system has been 
solved exactly for arbitrary initial conditions using these techniques.  The 
case of {\sf N} coupled oscillators in the presence of damping is also 
considered. It would be gratifying if these results can be experimentally verified.

\section{Acknowledgment}
One of us (VS) wishes to thank the Head of the Department of Theoretical Physics,
Professor R Chakrabarti for hospitality. VS and ASV also thank Professor M S Sriram
for discussions.

\end{document}